\documentclass[iop,apjl]{emulateapj}

\newcommand{\q}[1]{_{\rm #1}}
\newcommand{\ten}[1]{$10^{#1}$}
\newcommand{\scit}[2]{$#1\times10^{#2}$}
\newcommand{\scim}[2]{#1\times10^{#2}}

\newcommand{\ps}{s$^{-1}$}
\newcommand{\pcc}{cm$^{-3}$}
\newcommand{\smpy}{$M_\odot$ yr$^{-1}$}

\newcommand{\fig}[1]{Fig.\ \ref{fig:#1}}
\newcommand{\figg}[1]{Figure \ref{fig:#1}}

\newcommand{\sect}[1]{Sect.\ \ref{sec:#1}}

\newcommand{\mh}{H$_2$}
\newcommand{\mhm}{{\rm H}_2}
\newcommand{\hhhp}{H$_3^+$}
\newcommand{\hcop}{HCO$^+$}
\newcommand{\cdo}{CO$_2$}
\newcommand{\mn}{N$_2$}
\newcommand{\nnhp}{N$_2$H$^+$}
\newcommand{\amh}{NH$_3$}

\slugcomment{Accepted by ApJL June 26, 2012 -- erratum \today}

\shorttitle{Fundamental aspects of episodic accretion chemistry}
\shortauthors{Visser \& Bergin}

\begin{document}

\title{Fundamental aspects of episodic accretion chemistry explored with single-point models}

\author{Ruud Visser and Edwin A. Bergin}
\affil{Department of Astronomy, University of Michigan, 500 Church Street, Ann Arbor, MI 48109-1042, USA\\visserr@umich.edu, ebergin@umich.edu}

\begin{abstract}
We explore a set of single-point chemical models to study the fundamental chemical aspects of episodic accretion in low-mass embedded protostars. Our goal is twofold: (1) to understand how the repeated heating and cooling of the envelope affects the abundances of CO and related species; and (2) to identify chemical tracers that can be used as a novel probe of the timescales and other physical aspects of episodic accretion. We develop a set of single-point models that serve as a general prescription for how the chemical composition of a protostellar envelope is altered by episodic accretion. The main effect of each accretion burst is to drive CO ice off the grains in part of the envelope. The duration of the subsequent quiescent stage (before the next burst hits) is similar to or shorter than the freeze-out timescale of CO, allowing the chemical effects of a burst to linger long after the burst has ended. We predict that the resulting excess of gas-phase CO can be observed with single-dish or interferometer facilities as evidence of an accretion burst in the past \ten{3}--\ten{4} yr.
\end{abstract}



\section{Introduction}
\label{sec:intro}
The wide spread in bolometric luminosities observed among low-mass embedded protostars is at odds with a constant mass accretion rate onto the central source \citep{kenyon90a,kenyon94a,dunham08a,enoch09b,evans09a}. Several explanations have been put forward for this ``luminosity problem,'' the most common of which is episodic accretion: protostars accrete most of their mass in short bursts, while spending most of their time in a low-luminosity quiescent state \citep{kenyon90a}. The hypothesis of episodic accretion is supported by a suite of models \citep{young05a,vorobyov05b,vorobyov10b,zhu09a,zhu10b,dunham10a,dunham12a}, but many key questions remain unanswered. How intense are the accretion bursts? How many bursts does a given protostar experience, and on what timescales do they occur? How does episodic accretion affect the chemical evolution from a molecular cloud to a circumstellar disk? This Letter focuses on the question of episodic accretion chemistry and explores how one might use certain chemical tracers to help answer the questions outlined above.

The lack of constraints on most aspects of episodic accretion in embedded Class 0 and I protostars comes from the difficulty of observing the accretion bursts directly. In more evolved Class II sources, accretion bursts show up as FU Ori and EX Ori events \citep{herbig77a,hartmann96a}. Luminosity flares of factors of 2--10 have been observed for a few late Class I sources \citep{kospal07a,caratti11a}. Other evidence for variable accretion in embedded sources is all indirect, such as the luminosity problem \citep{kenyon90a} or the presence of periodic shocks along protostellar jets \citep{reipurth89a}. Additional probes are needed, and chemical signatures may be one.

To date, only two studies have addressed the topic of episodic accretion chemistry in embedded protostars. \citet{lee07a} showed that repeated burst--quiescent cycles lead to repeated freeze-out--evaporation cycles of CO and other species. The freeze-out timescale is longer than the duration of a burst, so the chemical effects of the burst can remain visible through part or all of the subsequent quiescent stage. \citet{kim11a} hypothesized that every time CO freezes out, a fraction is converted into \cdo{} ice. They found a better fit to the 15.2 \micron{} \cdo{} ice feature in the low-luminosity source CB130-1 -- presumed to be in a quiescent stage -- if 80\% of CO is converted into \cdo. It is unclear, however, why the conversion would have to happen in between accretion bursts rather than before the onset of collapse.

\citet{lee07a} and \citet{kim11a} used a full envelope model with fixed timescales for the burst and quiescent stages. In this Letter, we explore episodic accretion chemistry at an even more fundamental level by way of single-point models at various densities and timescales (Sects.\ \ref{sec:model} and \ref{sec:res}). We identify the key chemical processes affected by episodic accretion and illustrate how the physical and chemical timescales play against each other. We also offer some suggestions on how the chemical effects of episodic accretion may be observed with single-dish and interferometer facilities (\sect{disc}).


\section{Theory and model description}
\label{sec:model}
The fundamental aspects of the effects of episodic accretion on the chemistry of embedded protostars are best explored with a simple parametric model. Every accretion burst is accompanied by an increase in the protostellar luminosity, which raises the dust and gas temperatures throughout the envelope. Once the burst ends, the temperatures return to the original quiescent values. The chemical response to the temperature changes in the envelope depends on the duration of the bursts and quiescent periods and on the timescales of the dominant chemical reactions. The heating and cooling timescales for the gas and dust are less than a day \citep{draine03a} and do not feature into the discussion.

At typical envelope temperatures of at most 100 K, the chemistry following a decrease in temperature is dominated by the freeze-out of some gas-phase species onto the cold dust. Conversely, a heating event results in thermal evaporation of some ices. The evaporation rate depends exponentially on the dust temperature and is independent of the gas density; it is effectively instantaneous ($\tau\q{ev}<1$ yr) once the temperature exceeds a given species' evaporation temperature. The freeze-out rate is governed by the collision rate between molecules and grains \citep{charnley01a}. The freeze-out timescale of CO and \mn{} onto 0.1 \micron{} grains is
\begin{equation}
\label{eq:taufr}
\tau\q{fr} = \scim{1}{4}\,{\rm yr}\,\sqrt{\frac{10\,{\rm K}}{T\q{g}}} \frac{10^6\,{\rm cm}^{-3}}{n(\mhm)}\,,
\end{equation}
for a sticking coefficient of unity \citep{bisschop06a}.

The chemical timescales have to be balanced against the duration of the accretion bursts ($\tau\q{b}$) and of the quiescent periods between subsequent bursts ($\tau\q{q}$). In the models of \citet{zhu10b} and \citet{vorobyov10b}, bursts typically occur every \ten{3}--\ten{4} yr. The duration of each burst is at most a tenth of the duration of the quiescent phase. For some initial conditions, the models predict no substantial bursts at all.

Having identified the relevant timescales, the next question is how much the envelope is heated during a burst. At early times, appropriate for embedded protostars, the luminosity is dominated by accretion \citep{myers98a}. The models of \citet{zhu10b} and \citet{vorobyov10b} have mass accretion rates of $\dot{M}\q{q}=10^{-9}$--\ten{-7} \smpy{} in the quiescient phases and of $\dot{M}\q{b}=10^{-6}$--\ten{-4} \smpy{} during accretion bursts. For a stellar mass of 0.3 $M_\odot$ and a stellar radius of 2 $R_\odot$, the corresponding accretion luminosities ($L\q{acc} \approx G M_\ast \dot{M}/R_\ast$) are 0.005--0.5 and 5--500 $L_\odot$, respectively. We use the continuum radiative transfer code RADMC \citep{dullemond04a} to compute dust temperature profiles for an envelope exposed to central luminosities between 0.005 and 500 $L_\odot$. Our model envelope extends from 20 to \ten{4} AU and has a mass of 1 $M_\odot$, distributed along a power-law density profile with an exponent of $-1.5$. The gas-to-dust mass ratio is set to 100 and the dust opacities are taken from \citet{ossenkopf94a} for thin mantles at \ten{6} \pcc{} (OH5). The goal is to get a rough idea of how much the temperature increases during a burst, so the exact choice of envelope parameters and dust properties is not important.

\begin{figure}[t!]
\epsscale{1.12}
\plotone{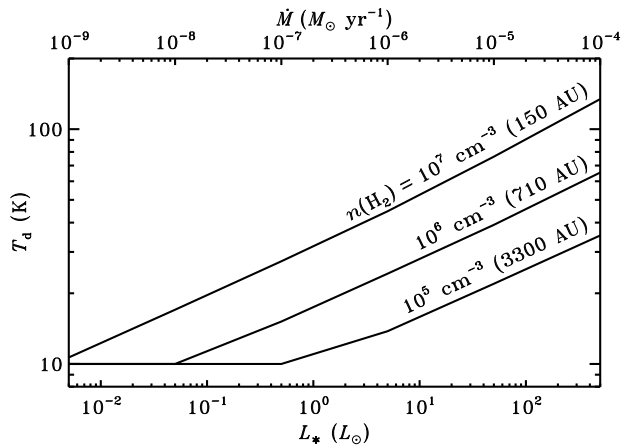}
\caption{Dust temperatures in a spherical envelope as function of the stellar luminosity and the mass accretion rate. The three curves correspond to the radii where the density reaches \ten{5}, \ten{6}, and \ten{7} \pcc.\label{fig:t567}}
\end{figure}

\figg{t567} shows the temperatures at the radii where the density is \ten{5}, \ten{6}, and \ten{7} \pcc. A lower limit of 10 K is applied in the low-luminosity cases to account for external heating by cosmic rays and the interstellar radiation field \citep{galli02a}. An accretion burst with an intensity of $I\q{b} \equiv \dot{M}\q{b}/\dot{M}\q{q}=100$ or 1000 increases the temperature by a factor of 2.5 or 4, respectively. This relationship holds regardless of the density power-law exponent and other envelope parameters.

We explore the chemical aspects of episodic accretion in a set of single-point models at densities of \ten{5}, \ten{6}, and \ten{7} \pcc. For each density, the dust and gas temperatures \citep[assumed coupled;][]{galli02a} are cycled between 10 and 30 K, corresponding to an accretion burst with an intensity of $I\q{b}=240$. The temperatures of 10 and 30 K fall below and above the CO and \mn{} evaporation temperatures of about 20 K \citep{bisschop06a}, so we expect the overall chemical effects to be dominated by their freeze-out--evaporation cycles.

We adopt the UDfA06 chemical network \citep{woodall07a}, expanded with freeze-out and thermal evaporation of all neutral species. The binding energies of CO, \mn, H$_2$O, \amh, and \cdo{} are set to 855, 800, 5773, 2790, and 2300 K \citep{fraser01a,bolina05a,bisschop06a,noble12a}. We follow the chemistry for five burst--quiescent cycles with $\tau\q{b}=100$ yr at 30 K (burst) and $\tau\q{q}=10^3$ or \ten{4} yr at 10 K (quiescent). The computation begins with typical dense pre-stellar core conditions as constrained by observations and models \citep{caselli99a,gibb04a,maret06a}: hydrogen in H$_2$; carbon in CO (80\%), CO ice (10\%), and \cdo{} ice (10\%); remaining oxygen in H$_2$O ice; and nitrogen in atomic N (60\%), \mn{} (20\%), \mn{} ice (10\%), and \amh{} ice (10\%). Also included are He, S, Si, Fe, Na, Mg, P, and Cl. Elemental abundances are taken from \citet{aikawa08a} and the cosmic-ray ionization rate of \mh{} is set to \scit{5}{-17} \ps{} \citep{dalgarno06a}.


\begin{figure*}[t!]
\epsscale{1.12}
\plotone{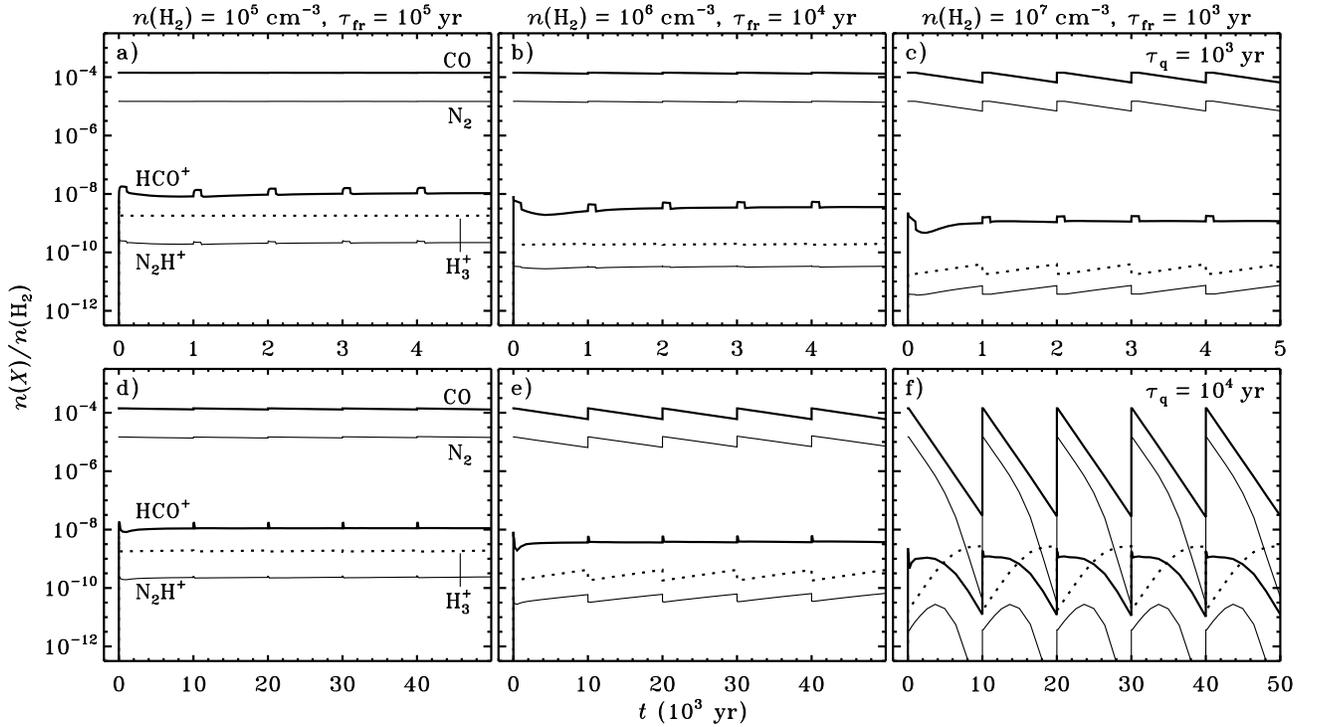}
\caption{Abundances of CO, \mn, \hcop, \nnhp, and \hhhp{} as function of time through five episodic accretion cycles in single-point models with bursts every \ten{3} or \ten{4} yr (top to bottom) and with densities of \ten{5}, \ten{6}, or \ten{7} \pcc{} (left to right). The dotted lines represent \hhhp; the other lines are as indicated in panels a and d.\label{fig:spabun}}
\end{figure*}

\section{Results}
\label{sec:res}
The temporal variation of the gas-phase abundances of CO, \mn, \hcop, \nnhp, and \hhhp{} is plotted in \fig{spabun}. The six panels show how the chemistry is controlled by the interplay of the dominant physical and chemical timescales. In panel f, the freeze-out timescale of CO and \mn{} is the same as the duration of the quiescent phase ($\tau\q{fr}=\tau\q{q}=10^4$ yr). Both species deplete by several orders of magnitude before the next burst hits. If we decrease the density by a factor of 10 or reduce the spacing between bursts by the same factor (panels c and e), freeze-out affects only half of the available CO and \mn. At even lower densities (panels a, b, and d), the freeze-out timescale is 100 or 1000 times longer than the duration of the quiescent phase, so CO and \mn{} remain fully in the gas phase.

The chemistry of \hcop, \nnhp, and \hhhp{} is closely tied to that of CO and \mn. \hhhp{} is destroyed mainly by reactions with CO and \mn{}, which act as the primary formation channels for \hcop{} and \nnhp. The primary destruction channels differ for \hcop{} and \nnhp: for the former it is dissociative recombination with electrons, producing CO and \mh, and for the latter it is proton exchange with CO, producing \hcop{} and \mn. All these reactions are fast, with timescales of about $0.1\,{\rm yr}\,[10^6\,{\rm cm}^{-3}/n(\mhm)]$ at the beginning of each quiescent phase. The episodic accretion chemistry of \hcop, \nnhp, and \hhhp{} is therefore controlled by the freeze-out timescales of CO and \mn.

\hcop, \nnhp, and \hhhp{} have the most complex abundance profiles in panel f of \fig{spabun}, where the dominant chemical timescale (freeze-out) is equal to the spacing between subsequent bursts. In each cold stage, \hhhp{} becomes more abundant as CO and \mn{} freeze out. The reverse happens during each burst: the rapid evaporation of CO and \mn{} results in a rapid loss of \hhhp. The abundance of \hcop{} remains constant between bursts as long as the decrease in CO is balanced by an increase in \hhhp. When the \hhhp{} abundance levels off, \hcop{} can no longer be formed as efficiently, and its abundance decreases. \hcop{} is reformed during the bursts from the evaporation of CO\@. The third species, \nnhp, shows a humpback pattern in every cold phase. This pattern results from the depletion of CO early on in the quiescent phase (slowing destruction of \nnhp) and the levelling off of \hhhp{} combined with ongoing depletion of \mn{} at later times (slowing formation of \nnhp).

If the model parameters are changed so that $\tau\q{fr}=10\tau\q{q}$ (panels c and e), only half of the gas-phase CO and \mn{} still freezes out in between bursts. The abundance patterns of \hhhp{} and \nnhp{} are now inversely proportional to those of CO and \mn, and \hcop{} remains constant except for small variations due to the inverse temperature dependence of dissociative recombination \citep{mitchell90a}. In panels a, b, and d, the dominant chemical timescales are so much longer than the physical timescales that all abundances remain constant in time except for the same small variations in \hcop.

The initial conditions have little effect on the above results. All CO and \mn{} ice evaporates in the first burst at $t=0$, so it does not matter what fraction of each species starts as gas or as ice. The ratios of CO versus \cdo{} and of \mn{} versus \amh{} affect the absolute abundances of CO and \mn, but leave the nature of the episodic abundance variations in \fig{spabun} unchanged.

Although these single-point models are only a crude analog to an actual protostellar envelope, they do highlight two fundamental aspects of episodic accretion chemistry: timescales matter, and the ``chemical temperature'' tends to be higher than the dust temperature. The first point is obvious from \fig{spabun}: the abundances only show strong variability if the duration between two accretion bursts is comparable to or exceeds the dominant chemical timescale, which in this case is the freeze-out timescale of CO and \mn. The second point is addressed in more detail in the next section.


\begin{figure*}[t!]
\epsscale{1.12}
\plotone{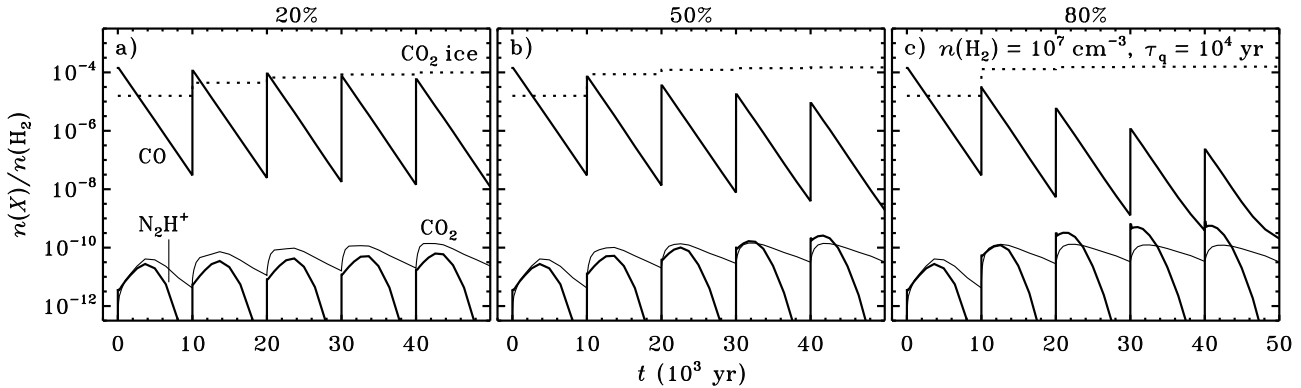}
\caption{Abundances of CO, \cdo, \cdo{} ice, and \nnhp{} as function of time through five episodic accretion cycles in single-point models with bursts every \ten{4} yr and a density of \ten{7} \pcc. The dotted lines represent \cdo{} ice; the other lines are as indicated in panel a. The panels differ in the fraction of CO ice converted into \cdo{} ice in each quiescent stage.\label{fig:co-co2}}
\end{figure*}

\section{Discussion}
\label{sec:disc}


\subsection{The chemical temperature and implications for observations}
\label{sec:tchem}
Whenever the stellar luminosity changes, the temperature of the gas and dust throughout the envelope responds on timescales of at most a day \citep{draine03a}. The chemistry responds equally fast to heating events, but it takes at least \ten{3} yr to adjust to a drop in temperature. This can lead to a prolonged situation where the dust and gas temperatures lie below the evaporation temperature of a given species, but the species remains in the gas phase because it has not had enough time yet to freeze out. Observationally, we therefore predict a mismatch between the temperature inferred from dust continuum emission and the ``chemical temperature'' inferred from the molecular abundances. In other words, the molecule does not exhibit the depleted abundance profile expected based on the dust temperature. \citet{aikawa07a} employed similar arguments to explain how vertical mixing in a circumstellar disk can sustain a reservoir of gas-phase CO below 20 K.

As an example, in panels a--c of \fig{spabun}, each model spends 90\% of the time at a temperature of 10 K and 10\% at 30 K\@. Observations of the dust emission would recover those fractions correctly. However, CO spends most of the time in the gas phase, so observations of the CO emission would imply a temperature of 30 K nearly 100\% of the time. In the absence of accretion bursts, this dense gas should be dominated by freeze-out. Hence, the mismatch in temperatures inferred from the dust and from the gas-phase abundances can be used as a novel probe for episodic accretion in embedded protostars.

The situation is more complicated in reality because a protostellar envelope spans a range of densities and temperatures: typically about \ten{4}--\ten{9} \pcc{} and 10--250 K \citep{jorgensen02a}. In a given accretion burst cycle, only a limited part of the envelope is heated from below to above the CO evaporation temperature. Nevertheless, our fundamental conclusion about relative timescales still applies: if the duration between bursts is shorter than the freeze-out timescale in the affected part of the envelope, CO remains in the gas phase \citep{lee07a}. Hence, if an envelope is observed between bursts, we expect to find an excess of gas-phase CO relative to the inferred dust temperature profile.

How could one probe such an excess? The CO abundance profile is commonly constrained by single-dish observations of pure rotational emission lines from the optically thin C$^{18}$O isotopolog \citep{jorgensen02a}. By combining spectra from a series of rotational lines with dust continuum data, it is possible in principle to infer the temperature at which CO freezes out. The excess CO gas exists at temperatures below the actual freeze-out temperature, so the temperature inferred from the observations should be lower than the value of about 20 K measured in the laboratory \citep{bisschop06a}. However, even with access to seven of the ten C$^{18}$O lines from $J=1$--0 to 10--9, the freeze-out temperature derived from observations is poorly constrained \citep{yildiz10a}. Single-dish observations of CO isotopologs in any one particular source are therefore unlikely to be a viable probe of episodic accretion chemistry.

We suggest two alternative observational probes of excess CO due to episodic accretion. The first is to take a large sample of protostars, such as that from the c2d \emph{Spitzer} legacy project \citep{evans09a}, and search for correlations between low-$J$ C$^{18}$O line intensities (directly tracing the column of cold CO) and systemic properties such as bolometric temperature, bolometric luminosity, and envelope mass. The second method is to directly measure the size of the C$^{18}$O emitting region in a sample of protostars using interferometers such as the SMA or ALMA. If there was a recent burst, the size of the CO emitting region should extend into dust with temperatures below the freeze-out temperature. Either method would need to be coupled to a parametric study of line radiative transfer models to understand in more detail how the low-$J$ C$^{18}$O emission -- spatially resolved and unresolved -- depends on variations in $T_{\rm bol}$, $L_{\rm bol}$, and $M_{\rm env}$, as well as on variations in the episodic accretion timescales.


\subsection{Grain-surface chemistry of CO}
\label{sec:grain}
\citet{kim11a} suggested that a fraction of the CO ice is converted into \cdo{} ice in every quiescent stage. \cdo{} has an evaporation temperature of 35--40 K \citep{noble12a}. If the conversion takes place in part of the envelope where the quiescent temperature lies below the CO evaporation temperature, it is unlikely to get warm enough during a burst for \cdo{} to evaporate (\fig{t567}). The presence of a pure \cdo{} ice feature at 15.2 \micron{} in low-luminosity sources may therefore indicate an accretion burst in the past \citep{lee07a}.

We reran the model for $n({\rm H}_2)=10^7$ \pcc{} and $\tau\q{q}=10^4$ yr (\fig{spabun}f), converting either 20, 50, or 80\% of CO ice into \cdo{} ice in each quiescent stage. The resulting abundances of CO, \cdo, \cdo{} ice, and \nnhp{} are plotted in \fig{co-co2}. Regardless of the conversion fraction, the abundance of \cdo{} ice rapidly goes up to about \ten{-4}. The gas-phase abundance fluctuates between \scit{3}{-11} and \scit{1}{-10} in response to the abundance variations of \hhhp{} (\fig{spabun}). \nnhp{} gradually becomes more abundant in each burst--quiescent cycle as its main destroyer, CO, is lost.

If a substantial amount of CO is indeed converted into \cdo, we would no longer expect to see an excess of gas-phase CO (\sect{tchem}), but instead we would expect a higher abundance of \nnhp{} between bursts. Conversion of CO ice into other species, such as H$_2$CO and CH$_3$OH, would have the same effect. At lower densities or a shorter quiescent stage, the amount of CO ice built up in each quiescent phase is smaller than for the case explored in \fig{co-co2}, and the effects of grain-surface conversion of CO into \cdo{} would be subdued. A grid of full envelope models is needed to explore quantitatively how much pure \cdo{} ice is formed as function of various parameters. As it stands, our single-point models suggest that the abundances of \cdo{} ice and gas-phase CO and \nnhp{} together are a novel probe to elucidate episodic accretion in low-mass embedded protostars.


\section{Conclusions}
\label{sec:conc}
Low-mass protostars appear to accrete most of their mass in a series of bursts interspersed with relatively long phases of very little accretion. If this hypothesis of episodic accretion is indeed true, the protostellar envelope is exposed to multiple heating and cooling events over the course of its existence. This Letter presents a fundamental theoretical investigation into how this would affect the chemical composition and evolution of the gas and dust in the envelope.

Based on simple single-point models, we conclude that the effects of episodic accretion chemistry depend strongly on the relative physical and chemical timescales. The abundances of common species like CO, \mn, \hcop, and \nnhp{} show strong time variability if the duration between subsequent accretion bursts is comparable to or longer than the dominant chemical timescale. For typical cold envelope conditions, the dominant chemical timescale is the freeze-out timescale of CO and \mn. Regardless of the exact timescales, the chemical effects of an accretion burst should remain visible for \ten{3}--\ten{4} yr after the burst has passed. In particular, we predict excess amounts of CO gas to be observable with either single-dish or interferometer facilities as chemical signposts of episodic accretion.


\acknowledgments
This work was supported by the National Science Foundation under grant 1008800.


\end{document}